\documentclass{article}
\usepackage{graphicx}
\usepackage{amsmath}
\usepackage{times}
\usepackage{url}
\usepackage{natbib}

\newtheorem{theorem}{Theorem}[section]
\newtheorem{lemma}{Lemma}[section]

\begin{document}
\title{ A consistent multivariate test of association based on ranks of distances}

\maketitle

\begin{center}
Ruth Heller \\
\emph{Department of Statistics and Operations Research, Tel-Aviv
university, Tel-Aviv, Israel. E-mail: ruheller@post.tau.ac.il}\\
Yair Heller \\
\emph{E-mail: heller.yair@gmail.com} \\
Malka Gorfine \\
\emph{Faculty of Industrial  Engineering and Management, Technion --
Israel Institute of Technology, Haifa, Israel.
E-mail:gorfinm@ie.technion.ac.il}
\end{center}

\begin{abstract}
We are concerned with the detection of associations  between random
vectors of any dimension. Few tests of independence exist that are
consistent against all dependent alternatives. We propose a powerful
test that is applicable in all dimensions and is consistent against
all alternatives. The test has a simple form and is easy to
implement. We demonstrate its good power properties in simulations
and on examples.
\end{abstract}

\section{Introduction}
In modern applications, there is need to test for independence
between random vectors. One example from genomics research is
whether two groups of genes are associated. Another application is
functional magnetic resonance imaging  research, where voxels in the
brain are measured over time under various experimental conditions,
and it is of interest to discover whether sets of voxels that
comprise different areas in the brain are functionally related.

Let ${ X}\in \Re^p$ and ${Y} \in \Re^q$ be random vectors, where $p$
and $q$ are positive integers. We are interested in testing whether
there is a relationship between  the two vectors $ X$ and $ Y$. The
null hypothesis states that the two vectors are independent,
$$H_0: F_{XY} =F_X F_Y,$$
where the joint distribution of $({ X},{ Y})$ is denoted by
$F_{XY}$, and the distributions of ${ X}$ and ${ Y}$, respectively,
by $F_X$ and $F_Y$. We are interested in the general alternative
that the vectors are dependent, $$H_1: F_{XY} \neq F_X F_Y.$$ There
are $N$ independent copies $({ x}_i,{ y}_i)$, $i=1,\ldots,N$ from
the joint distribution of $ X$ and $ Y$ for testing $H_0$. The
dimensions of the vectors $p$ and $q$ may be much higher than $N$.

The purpose of this paper is to provide a powerful test of
independence that is applicable in all dimensions,  and is
consistent against all alternatives. The test is based on the
pairwise distances between the sample values of $ X$ and of $ Y$
respectively, $\{d_X({  x_i}, {  x_j}): i,j \in \{1,\ldots,N \} \}$,
$\{d_Y({  y_i}, {  y_j}): i,j \in \{1,\ldots,N \} \}$. The only
restriction on the distance metrics $d_X(\cdot,\cdot)$ and
$d_Y(\cdot, \cdot)$ is that they are determined by norms. The test
statistic is a function of ranks of these distances, and it can be
expressed simply in closed form. It is proven to be consistent
against all dependent
alternatives. %It is shown in simulations to
%have excellent power against non-monotone alternatives.

Few multivariate tests of independence that are consistent against
all alternatives are available to date. \cite{fukumizu} suggest a
test based on normalized cross-covariance operators on reproducing
kernel Hilbert spaces. \cite{Bickel09} offer a test based on an
approximation of Renyi correlation, since there is no explicit
formula to compute the Renyi correlation. A very elegant test with a
simple formula is provided in \cite{Szekely07}, and has been further
investigated in \cite{Szekely09} and in the discussions that
followed it. We revisit some of the examples of \cite{Szekely07},
and add new examples. In the examples considered our new test
performs remarkably well in comparison to the test of
\cite{Szekely07}.

\section{The new test of independence}\label{sec-newtest}
This section develops the new test of independence. To motivate the
test, note that if $  X$ and $  Y$ are dependent and have a
continuous joint density, then there exists a point $({  x}_0, {
y}_0)$ in the sample space of $({  X}, {  Y})$, and radii $R_x$ and
$R_y$ around ${  x}_0$ and ${  y}_0$, respectively, such that the
joint distribution of $  X$ and $  Y$ is different than the product
of the marginal distributions in the cartesian product of balls
around $({  x}_0, {  y}_0)$. Consider first an oracle that guesses
such a point
 $({  x}_0, {  y}_0)$  and radii $R_x$ and $R_y$. %in the sample space, of
%radius $Rx$ in $  X$ and of radius $Ry$ in $  Y$, the joint
%distribution of $  X$ and $  Y$ is very different than the
%marginal distributions.

Let $d(\cdot,\cdot)$ be the norm distance between two sample points,
either in $  X$ or in $  Y$, so the distance between the vectors ${
x}_i$ and ${  x}_j$ from the distribution of $  X$ is $d({  x}_i, {
x}_j)$, and similarly the distance between the vectors ${  y}_i$ and
${  y}_j$ from the distribution of $  Y$ is $d({  y}_i, { y}_j)$.
Technically, this distance may be different for the samples of $ X$
and for the samples of $  Y$, but we omit this distinction for
simplicity of notation. Consider the following two dichotomous
random variables: $I\{d({  x}_0,{  X})\leq R_x\}$ and $I\{d({ y}_0,{
Y})\leq R_y\}$, where $I(\cdot)$ is the indicator function. We
summarize the observed cross-classification of these two dichotomous
random variables for the $N$ independent observations $k\in
\{1,\ldots,N\}$ in Table \ref{tab-cont1}, where $A_{11} =
\sum_{k=1}^N I\{d({  x}_0,{  x}_k)\leq R_x\} I\{d({  y}_0,{
y}_k)\leq R_y\}$,  $A_{12}, A_{21}, A_{22},$ defined similarly, and
$A_{m \cdot}$, $A_{\cdot m} \quad m=1,2$, are the sum of the row or
column, respectively.

\begin{table}
\caption{The  cross-classification of $I\{d({ x}_0,{  X})\leq R_x\}$
and $I\{d({  y}_0,{  Y})\leq R_y\}$}\label{tab-cont1}
\begin{tabular}{c|c|c|c}
& $d({  y}_0,\cdot)\leq R_y$ &  $d({  y}_0,\cdot)> R_y$& \\
\hline $d({  x}_0,\cdot)\leq R_x$ & $A_{11}$ & $A_{12}$&
$A_{1\cdot}$
\\ \hline $d({  x}_0,\cdot)> R_x
$ & $A_{21}$ & $A_{22}$ & $A_{2\cdot}$\\
\hline & $A_{\cdot1}$ & $A_{\cdot2}$ & $N$
\end{tabular}
\end{table}

%If the joint distribution  is very different than the product of the
%marginal distributions within the cross of balls of radii $R_x$ and
%$R_y$ around $({  x}_0, {  y}_0)$, then $A_{11}$ will be very
%different than $A_{1\cdot}A_{\cdot 1}/N$. Therefore,
Evidence against independence may be quantified by Pearson's
chi-square test statistic, or the likelihood ratio test statistic,
for $2\times 2$ contingency tables. The test based on such a
statistic is consistent, and its power for finite sample size
depends on the choice of $({  x}_0, {  y}_0)$, $R_x$ and $R_y$.

Since we do not have an oracle that guesses well $({  x}_0, {
y}_0)$, $R_x$ and $R_y$, in the sense that the test for independence
by a $2\times 2$ contingency tables will be powerful, we let the
data guide us in these choices. For every sample point $i$, we
choose it in its turn to be $({  x}_0, {  y}_0)$. For every sample
point $j\neq i$, we choose it in its turn to define $R_x = d({ x}_i,
{  x}_j)$ and $R_y =d({  y}_i, {  y}_j)$. The $2\times 2$ tables now
comprise the remaining $N-2$ points. The test  aggregates the
evidence against independence by summing over all $N(N-1)$ test
statistics from the $2\times 2$ tables thus created.

Specifically, for fixed observations $i$ and $j$, consider the
 dichotomous random variables: $I\{d({  x}_i,{
X})\leq d({  x}_i,{  x}_j)\}$ and $I\{d({  y}_i,{  Y})\leq d({
y}_i,{ y}_j)\}$. Table \ref{tab-cont2} summarizes the observed
cross-classification of these two dichotomous random variables for
the $N-2$ independent observations $k\in \{1,\ldots,N\}, k\neq i,
k\neq j$, where  $A_{11}(i,j) = \sum_{k=1, k\neq i, k\neq j}^N
I\{d({ x}_i,{  x}_k)\leq d({  x}_i,{  x}_j)\}I(d({  y}_i,{ y}_k)\leq
d\{{ y}_i,{  y}_j)\}$, $A_{12}, A_{21}, A_{22}$ defined similarly,
and $A_{m \cdot}$, $A_{\cdot m},  m=1,2$, are the sum of the row or
column, respectively.

\begin{table}
\caption{The  cross-classification of $I\{d({ x}_i,{ X})\leq
d(x_i,x_j)\}$ and $I\{d({  y}_i,{  Y})\leq
d(y_i,y_j)\}$}\label{tab-cont2}
\begin{tabular}{c|c|c|c}
& $d(y_i,\cdot)\leq d(y_i,y_j)$ &  $d(y_i,\cdot)> d(y_i,y_j)$& \\
\hline $d(x_i,\cdot)\leq d(x_i,x_j)$ & $A_{11}(i,j)$ &
$A_{12}(i,j)$& $A_{1\cdot}(i,j)$ \\ \hline $d(x_i,\cdot)>
d(x_i,x_j)$ & $A_{21}(i,j)$ & $A_{22}(i,j)$ & $A_{2\cdot}(i,j)$\\
\hline & $A_{\cdot1}(i,j)$ & $A_{\cdot2}(i,j)$ & $N-2$
\end{tabular}
\end{table}

Let $$S(i,j) =
\frac{(N-2)\{A_{12}(i,j)A_{21}(i,j)-A_{11}(i,j)A_{22}(i,j)\}^2}{A_{1\cdot}(i,j)A_{2\cdot}(i,j)A_{\cdot1}(i,j)A_{\cdot2}(i,j)
}.$$ This is the classic test statistic for Pearson's chi square
test for $2\times 2$ contingency tables.

To test for independence between the two random vectors $  X$ and $
Y$, we suggest as a test statistic $T = \sum_{i=1}^N
\sum^N_{\substack{
 j=1 \\ j\neq i} }S(i,j). $
 For $i$ and $j$ with 0 in at least one of the margins, we set $S(i,j)=0$.
 The $p$-value from the permutation test based on the statistic $T$ is the fraction
 of replicates of $T$ under random permutations of the indices of
 the $  Y$ sample, that are at least as large as the observed
 statistic.

We say a point $(x_0, y_0)$ is a point of dependence if the  joint
density of $X$ and $Y$ is different than the product of the marginal
densities of $X$ and $Y$ at $(x_0,y_0)$, defined formally in
equation (\ref{app-eqnPointPositiveDependence}) in the Appendix for
the mixed case where the coordinates may be both discrete and
continuous. Theorem \ref{thm-consistenyGeneral} states that the test
is consistent for
 discrete random vectors with countable support, as well as for continuous random vectors, and for random vectors where some of the
coordinates are discrete and others continuous, if the density of
the continuous random vectors is continuous around a point of
dependence.

\begin{theorem}\label{thm-consistenyGeneral}
For dependent random vectors $( X,  Y)$, $X \in \Re^p$ and $Y\in
\Re^q$, denote the discrete and continuous coordinates of $ X$ by
$u\subseteq \{1,\ldots,p \}$ and $ v =  u^c$, respectively, and
similarly the discrete and continuous coordinates of $Y$ by $ s
\subseteq \{1,\ldots,q \}$ and  $ t =  s ^c$, respectively. The
permutation test based on the statistic $T$, with distances $d_{
X}(\cdot,\cdot)$ and $d_{  Y}(\cdot,\cdot)$ determined by norms, is
consistent if either
\begin{enumerate}
\item $X$ and $Y$ are continuous, i.e. $u$ and $s$ are empty sets, and   there exists a  point of dependence $(x_0,y_0)$ for which the joint density  is
continuous.
\item At least one of $X$ or $Y$ has discrete coordinates in
addition to the continuous coordinates, i.e. at least one of $u$ and
$s$ is non-empty and both $v$ and $t$ are non-empty, and there
exists a point of dependence $(x_0,y_0)$ for which (i) there exists
a ball around the atom $\{x_0(u), y_0(s)\}$ that contains only this
atom,  and (ii) the joint density of the continuous coordinates
conditional on the discrete coordinates is continuous.
\item Both $X$ and $Y$ are discrete, i.e. $v$ and $t$ are empty
sets.
\item $X$ is discrete and $Y$ is continuous, i.e. $v$ and $s$ are
empty sets, and there exists a point of dependence $(x_0,y_0)$ for
which the conditional density of $Y$ given $X$ is continuous.
\end{enumerate}
\end{theorem}
 See Appendix for a proof of case 2. The proofs of
 the other cases are very similar yet simpler,  and they are given in the Supplementary Material.

\subsection{Computational Complexity}\label{subsec-comp}

For $N$ sample points, the naive implementation of the test will
require an order of magnitude of $N^3$ operations. We provide an
algorithm to efficiently calculate the score $T$ in order of
magnitude $N^2\log N$. This is done by providing an algorithm which
for a given $i$ calculates $\{S(i,j): j=1,\ldots,N, j\neq i\}$
 in order of magnitude $N \log N$. We shall show that we can calculate $\{A_{11}(i,j), A_{12}(i,j),
A_{21}(i,j), A_{22}(i,j): j=1,\ldots,N, j\neq i\} $
 in $O(N\log N)$.

For fixed $i$, let us look at all the distances from sample $i$
according to $  X$ and let us sort the samples according to
distance. Without loss of generality, renumber the indices of the
$N-1$ sample points other than $i$ to be $1,\ldots,N-1$, so that the
$j$th observation is the $j$th nearest to $i$ in $  X$. Denote the
order of the distance from $i$ in $  Y$ by $\pi(1) \cdots \pi(N-1)$.
So the $j$th observation is the $\pi(j)$th nearest to $i$ in $ Y$.
$\pi(\cdot)$ is a permutation of $1,\ldots,N-1$. The entries in the
above Table \ref{tab-cont2} may be expressed as a function of $j$,
$\pi(j)$ and $inv(j)$, where $inv(j)$ is defined as the number of
inversions of $j$ in the permutation $\pi$, i.e. $inv(j)$ is the
number indices $k\in \{1,\ldots,j-1\}$ such that $\pi(k)\in
\{\pi(j)+1,\ldots,N-1\}$. From the definition of $A_{12}(i,j)$ it
follows that $A_{12}(i,j)=inv(j)$, and similarly
$A_{22}(i,j)=N-\pi(j)-inv(j)$. Since $A_{1\cdot}(i,j)=j-1$, the
remaining counts of  the $2\times 2$ contingency table for $S(i,j)$
are $A_{11} =j-1-inv(j), A_{21} = \pi(j)+inv(j)-j-1$. Therefore, it
is enough to show that each of the following steps takes order of
magnitude $N \log N$: (1) renumber the indices according to
increasing distance in $ X$ from $i$; (2) compute $\{ \pi(j):
j=1,\ldots,N, j\neq i \}$; (3) compute $\{ inv(j): j=1,\ldots,N,
j\neq i \}$. Since sorting takes order of magnitude $N \log N$,
steps (1) and (2) are performed in the required computational time.
It remains to show that (3) can be computed in order of magnitude $N
\log N$. We show the algorithm in the Supplementary Material.

\section{Simulations}\label{sec-sim}
% ADD MIC?

In the simulations, we compare the performance of our test and the
dCov test of \cite{Szekely09}. We chose the latter test  for two
reasons. First, it is the only
consistent test of simple form that is available. %, and it is
%similar to our test in the sense that it is based on the pairwise
%distance matrices of sample point in $  X$ and in $  Y$.
Second, the superiority of the dCov test over classical tests in
\cite{Puri71} has been demonstrated in \cite{Szekely07}. Moreover,
our aim is to investigate the performance of our test for
non-monotone relationships,  and these classical tests, or related
tests for higher dimensions found in \cite{Taskinen05}, are
ineffective for testing non-monotone types of dependence
\citep{Szekely07}.

In all simulations, the dCov test  was applied by calling the
function $dcov.test$ implemented in the R package \emph{energy }
\citep{Szekely09} with 10000 permutation samples. The Euclidean
distance was used as a distance metric.

We consider first the six simulated examples of unusual bivariate
distributions in \cite{Newton09}. These examples mimic those at the
\url{wikipedia.org} page on Pearson correlation, see Supplementary Material for details.   %For $N=100$
%sample points the relations are already manifest by eye, as can be
%seen from the example data in Figure \ref{figNewton100}.
The example of 4 independent clouds is an example of a null
distribution. Table \ref{tab1} shows the power comparison between
 dCov and the new test for $N=50$ sample points and a significance level $\alpha = 0.05$.   Large differences
are observed. The most pronounced difference is observed for the
circle relation, where the power of the new test is 0.993 yet dCov
has no power to detect the relation. For the diamond relation, the
new test has a power of 0.662 yet the power of dCov is 0.037. The
tests based on Pearson and Spearman correlations had a power of at
most 0.16 in all examples.

\begin{table}[!t]
\caption{The power ($SE\times 100$) for a test at level 0.05 from a
sample of size $N=50$ from unusual bivariate relations. The results
are based on 1000 simulations for rows $1-5$ and on 50000
simulations for the null setting in row $6$.}\label{tab1}
\begin{center}
\begin{tabular}{c|c|c|}
  Distribution           &     Dcov  &       new test \\ \hline
 W & 0.853 (1.1) & 1.000 (0.0) \\
 Diamond & 0.037 (0.3) & 0.662 (1.5) \\
 Parabola & 0.975 (0.5) & 0.998 (0.1) \\
 2 Parabolas & 0.303 (1.4) & 1.000 (0.0) \\
 Circle & 0.000 (0.0) & 0.993 (0.3) \\
 4 independent clouds & 0.050 (0.1) & 0.050 (0.1)\\
\end{tabular}
\end{center}
\end{table}

\cite{Szekely07} considered multivariate examples and compared them
to likelihood ratio type of tests. In the following two examples
from \cite{Szekely07}, none of the likelihood ratio type of tests
considered performed well. Using our notation, the distribution of
${  X}= (X_1,\ldots,X_5)$ is standard multivariate normal with 5
dimensions. First, let $  Y$ be equal to $\log({  X^2})$. Columns 2
and 3 of Table \ref{tab-szekely07ex3} shows the power of a test at
level $0.05$ for dCov as well as for the new test.  The new test has
a power of 0.82 for $N=40$ sample points, whereas the power of dCov
is 0.436. Second, let ${  Y}= (Y_1,\ldots,Y_5)$ have coordinates
$Y_j = X_j\cdot \epsilon_j$, where $\epsilon_j$ are independent
standard normal variables and independent of $  X_j$. Columns 4 and
5 of Table \ref{tab-szekely07ex3} show the power of a test at level
$0.05$ for dCov as well as for the new test.  The new test has a
power of 0.968 for $N=50$ sample points, whereas the power of dCov
is 0.443.

\begin{table}[!t]
\caption{The power ($SE\times 100$) of a test at level $0.05$ per
sample size from a 5 dimensional joint distribution, where $  X\sim
N(0, I_{5\times 5})$  and ${  Y} = \log({ X}^2)$ or
 ${  Y}= (Y_1,\ldots,Y_5)$ has coordinates $Y_j =
X_j\cdot \epsilon_j$, where $\epsilon_j\sim N(0, 1)$ independent of
$ X_j $. The results are based on 1000
simulations.}\label{tab-szekely07ex3}
\begin{center}
\begin{tabular}{r|rr|rr}
  & \multicolumn{2}{|c|}{${  Y} = \log({  X}^2)$ } & \multicolumn{2}{|c}{ $Y_j = X_j\cdot \epsilon_j$
  }\\
 Sample size & dCov & new test &  dCov & new test  \\ \hline
N=20 & 0.172 (1.2) & 0.299 (1.4) & 0.335 (1.5) & 0.554 (1.6)\\
N=30 & 0.290 (1.4) & 0.595 (1.6) & 0.384 (1.5) & 0.792 (1.3)\\
N=40 & 0.436 (1.6)& 0.819 (1.2) & 0.417 (1.6) & 0.920 (0.9)\\
N=50  & 0.629 (1.5)& 0.945 (0.7) & 0.443 (1.6) & 0.968 (0.6)\\
\end{tabular}
\end{center}
\end{table}

A more sophisticated scenario, which includes both a monotone and
non-monotone component, is the following: $Y_j = \beta_1 X_j+\beta_2
X_j^2 + \epsilon_j, j=1,\ldots,m_1$ and $Y_j = \epsilon_j,
j=m_1+1,\ldots, 5$, with $\epsilon_j \sim N(0, \sigma^2)$ and
$X_j\sim N(0,1)$ for all $j$. Table \ref{tab-referee} shows the
power of a test at level $0.05$ for dCov as well as for the new test
for various values of $\beta_1, \beta_2, \sigma^2$, $m_1 \in
\{0,2\}$. Further results  in $100$ dimensions are included in the
Supplementary Material. When $\beta_2$ is large relative to
$\beta_1$, the power of the new test is better than that of dCov.

\begin{table}[!t]
\caption{The power ($SE\times 100$) of a test at level $0.05$ per
sample size from a 5 dimensional joint distribution, where $Y_j =
\beta_1 X_j+\beta_2 X_j^2 + \epsilon_j, j=1,\ldots,m_1$ and $Y_j =
\epsilon_j, j=m_1+1,\ldots, 5$, with $\epsilon_j \sim N(0,
\sigma^2)$ independent of $X_j\sim N(0,1)$. The results are based on
1000 simulations.}\label{tab-referee}
\begin{center}
\begin{tabular}{rrrr|rr|rr}
   &&&& \multicolumn{2}{|c|}{dCov } & \multicolumn{2}{|c}{new test
  }\\
$m_1$ & $\beta_1$ & $\beta_2$ & $\sigma^2$ & N=20 & N=30 &  N=20 & N=30  \\
\hline
0 & 0 & 0 & 1 & 0.040 (0.6) & 0.047 (0.7)  & 0.051 (0.7) & 0.047 (0.7) \\
2 & 1 & 4 & 9 & 0.501 (1.6) & 0.637 (1.5) & 0.669 (1.5) & 0.984 (0.4)\\
2 & 3 & 2.5 & 9 & 0.841 (1.2) & 0.963 (0.6) & 0.706 (0.5) & 0.998 (0.1)\\
 \hline
\end{tabular}
\end{center}
\end{table}

Finally, we consider an example where $X$ and $Y$ are both of
dimension 1000, from a mixture distribution with 10 equally likely
components. In the $i$th component, $i\in \{1,\ldots,10\}$, $(X,Y)$
are the random variables $\{\mu_x(i) +\epsilon, \mu_y(i)+\eta\}$,
where
 $\mu_x(i)$ and $\mu_y(i)$
are sampled (once) from the 1000 dimensional multivariate standard
normal distribution, and $(\epsilon, \eta)$ are sampled
independently from the multivariate Cauchy or multivariate $t$ with
3 degrees of freedom, with the identity correlation matrix. The
dependency of $X$ and $Y$ is through the fixed pairs $\{\mu_x(i),
\mu_y(i)\}, i=1,\ldots,10$ such that the data consists of 10 clouds
around these pairs. See Supplementary Material for details. Table
\ref{tab-qpgeqn} shows the power of a test at level $0.05$ for dCov
as well as for the new test.  The new test has a power of one for
$N=200$ sample points in the multivariate $t$ distribution, whereas
the power of dCov is 0.23. For the multivariate cauchy distribution,
dCov has no power even at $N=300$, as expected since dCov is
consistent only for distributions with finite first moments
\citep{Szekely07}. The power of the new test is 0.58 for $N=300$
sample points. Moreover, for the multivariate normal distribution,
the power for both tests is one for $N=50$ sample points.

\begin{table}[!t]
\caption{The power ($SE\times 100$) of a test at level $0.05$ per
sample size from the joint distribution  of 10 mixture components
for random vectors of dimension 1000, each component is centered
around a different mean and  is either multivariate Cauchy or
multivariate t with 3 degrees of freedom. The results are based on
200 simulations.}\label{tab-qpgeqn}
\begin{center}
\begin{tabular}{r|rr|rr}

  & \multicolumn{2}{|c|}{t (3df)} & \multicolumn{2}{|c}{Cauchy }\\
 Sample size & dCov & new test & dCov & newtest \\
 \hline
 N=50 & 0.100 (2.1) & 0.570 (3.5)  & 0.040 (1.4) &  0.130 (2.4)\\
  N=100  & 0.190 (2.8) & 0.980 (1.0)  & 0.050 (1.5) & 0.185 (2.7) \\
  N=200 & 0.345 (3.4) & 1.000 (0.0)  & 0.075 (1.9) & 0.390 (3.5) \\
 N=300 & 0.620 (3.2) & 1.000 (0.0)  & 0.020 (1.0) & 0.580 (3.5)\\
  \hline
\end{tabular}
\end{center}
\end{table}

\section{An example}\label{sec-example}
In a homogeneous population, the dependence between single
nucleotide polymorphysms (SNPs) on the same chromosome is weaker the
farther the SNPs are from each other due to recombination
\citep{Lander94}. A question of interest is whether SNPs across
chromosomes are  independent. To answer this question we examined
the DNA of a sample of 97 unrelated individuals of Han Chinese in
Beijing, China, available from the HapMap project \citep{HapMap03}.
This sample is regarded to be of relatively homogeneous ancestry,
since donors were required to have at least three  Han Chinese
grandparents.  For the purpose of this example, we limit ourselves
to chromosomes 21 and 22 and ask whether the SNPs on chromosome 21
are independent of the SNPs on chromosome 22. We first preprocessed
the data by removing subjects with more than 30\% missing SNPs on a
chromosome, SNPs with missing subjects, and SNPs with minor allele
frequency below 0.05. After preprocessing, 43 subjects remained. For
each subject we had a vector of dimension 31,858 of SNPs from
chromosome 21, and a vector of dimension 36,264 of SNPs from
chromosome 22. The Euclidean distance was used as a distance metric.
Our proposed test was highly significant, with a $p$-value below
$1\times 10^{-4}$. The dCov test was also
significant, with a $p$-value of $6\times 10^{-4}$. %This example is
%crude, but it serves to highlight the potential of our test.
%Specifically, it may be useful in identifying whether a population
%is indeed of common ancestry.

\section{Final remarks}\label{sec-discuss}
 Pearson's chi-squared test
statistic was originally proposed as an approximation to the
log-likelihood ratio statistic, in our context
$$S_{LR}(i,j) = 2\sum_{k=1}^2\sum_{l=1}^2A_{kl}(i,j) \log
[A_{kl}(i,j)/\{\frac{A_{\cdot l}(i,j)A_{k\cdot}(i,j)}{N-2}\}].$$ An
alternative test statistic for independence may therefore be $T_{LR}
= \sum_{i=1}^N \sum^N_{\substack{ j=1 \\ j\neq i} }S_{LR}(i,j). $ In
the simulation results considered, the permutation test with this
test statistic had very similar power to the power of the suggested
test.

After discovering that the random vectors are dependent, a natural
question to ask is which sub-vectors are dependent. This can be done
using multiple comparisons procedures, similar to post-hoc testing
in the analysis of variance \citep{Scheffe59}. Moreover, the larger
the value of $S(i,j)$, the stronger the dependence between the
variables $I\{d({  x}_i,{  X})\leq d({  x}_i,{  x}_j)\}$ and $I\{d({
y}_i,{ Y})\leq d({  y}_i,{  y}_j)\}$. Informally, if $S(i,j)$ is
large and $d({  x}_i,{  x}_j)$ and $d({  y}_i,{ y}_j)$ are small,
this suggests that the random vectors $  X$ and $  Y$ are dependent
in balls of size $d({  x}_i,{  x}_j)$ and $d({ y}_i,{ y}_j)$ around
${ x}_i$ and ${  y}_i$. We plan to explore methods of localizing the
dependency in future work.

\section*{Acknowledgement}
We thank Noam Berger for very useful discussions of the theorem, and
David Golan for suggesting the example. Thank you also to the
editor, associate editor, and the two reviewers for the helpful
comments that led to substantial improvements to the manuscript.

\section*{Supplementary material}
Supplementary material includes the proofs of cases 3 and 4 of the
theorem, the algorithm for implementing the test in order of
magnitude $N^2\log(N)$, further simulations, and an additional
one-dimensional real data  example.

\appendix
\section*{Appendix} \label{appProof}

We shall prove Theorem \ref{thm-consistenyGeneral} for the case
where the index sets $u, v, s, t$ are all non-empty, since it is
straightforward to adapt the proof to the cases where $u$ or $s$ are
empty sets.

From henceforth, for notational convenience we shall repress the
conditioning event and denote the joint and marginal densities
conditional on the discrete coordinate values as  $h\{{  x}({  v}),
{ y}({  t})\}, f\{{  x}({  v})\}$, and  $g\{{  y}({  t})\}$ in place
of $h\{{  x}({  v}), {  y}({  t}) \mid {  X}({  u}) = {  x}({ u}), {
Y}({  s})  = {  y}({  s})\}, f\{{  x}({  v}) \mid {  X}({  u}) = {
x}({ u})\} $, and $g\{{  y}({  t}) \mid {  Y}({  s}) = {  y}({
s})\}$. Moreover, we denote $p\{{  x}({  u}), {  y}({  s})\} =Pr\{{
X}({ u}) = { x}({ u}), {  Y}({  s}) = {  y}({  s})\}, p\{{  x}({
u})\} = Pr\{{ X}({ u}) = {  x}({  u})\}$, and $p\{{  y}({  s})\} =
Pr\{{ Y}({ s}) = { y}({ s})\}$.

If $H_0$ is false,  and the point of dependence $(x_0, y_0)$
satisfies properties (i) and (ii) of Theorem
\ref{thm-consistenyGeneral}. Without loss of generality, suppose
\begin{equation}\label{app-eqnPointPositiveDependence}
p\{x_0(u), y_0(s)\}h\{x_0(v),y_0(t)\}>
p\{x_0(u)\}f\{x_0(v)\}p\{y_0(s)\}g\{y_0(t)\}.
\end{equation}

 Let  $R_d$ be a positive constant smaller than both the radius
of the ball around $x_0(u)$ that contains only $x_0(u)$, and  the
radius of the ball around $y_0(s)$ that contains only the point
$y_0(s)$. Then the set $\{(x,y): d(x,x_0)<R_d, d(y,y_0)<R_d\}$
contains only points with discrete coordinates $x(u) = x_0(u), y(s)
= y_0(s)$. Moreover, since the joint density conditional on
$\{x_0(u), y_0(s)\}$ is continuous, there exists a radius $R_c$ such
that $p\{x_0(u), y_0(s)\}h\{x(v),y(t)\}>
p\{x_0(u)\}f\{x(v)\}p\{y_0(s)\}g\{y(t)\}$ for all points $(x,y)$ in
the set $ \{(x,y): d(x,x_0)<R_c, d(y,y_0)<R_c, x(u) = x_0(u), y(s) =
y_0(s) \}$. Let $R = \min \{R_d, R_c\}$ and  $\mathcal{A} = \{(x,y):
d(x,x_0)<R, d(y,y_0)<R \}$. Then the set $\mathcal{A}$ has positive
probability, for all points $(x,y)\in \mathcal{A}$ the discrete
coordinates are $x(u) = x_0(u)$ and $y(s) = y_0(s)$, and moreover
$$\min_{\mathcal{A}}[p\{x(u), y(s)\}h\{x(v),y(t)\}
-p\{x(u)\}f\{x(v)\}p\{y(s)\}g\{y(t)\}  ]>0. $$  Denote this minimum
by the positive constant $c$.

Clearly the following two subsets of $\mathcal{A}$ have positive
probability as well: $$\mathcal{A}_1 = \{({  x}, {  y}): d({ x},{
x}_0)< R/8, d({  y},{  y}_0)<R/8 \}$$ and
$$\mathcal{A}_2 = \{({  x}, {  y}): 3R/8< d({
x},{  x}_0)< R/2, 3R/8<d({  y},{  y}_0)<R/2 \}.$$ Denote the
probabilities of $\mathcal{A}_1$ and $\mathcal{A}_2$ by $f_1$ and
$f_2$ respectively.
%we expect a positive fraction $f_1$ of samples to fall in
%the $  x$ and $  y$ spheres of radius $R/8$, and a positive
%fraction $f_2$ of samples to fall in the $  x$ and $  y$ 'tubes'
%a distance from $3R/8$ to $R/2$ around ${  x}_0, {  y}_0$:
%\begin{eqnarray}
%&& f_1 = \int_{\substack{\{({  x},{  y} ): d({  x},{  x}_0)<R/8\\
%d({  y},{  y}_0)<R/8\} }}h({  x},{  y})d{  x}d{  y} >0
%\\
%&&
%f_2 = \int_{\substack{\{({  x},{  y} ):3R/8<d({  x},{  x}_0)<R/2\\
%3R/8<d({  y},{  y}_0)<R/2\}}}h({  x},{  y})d{  x}d{
%y}>0
%\end{eqnarray}
Therefore, we expect $(N f_1) (N f_2)$ pairs of sample points $i$
and $j$ such that $({  x}_i, {  y}_i) \in \mathcal{A}_1$ and $({
x}_j, {  y}_j) \in \mathcal{A}_2$.
%$d({  x}_i, {  x}_0)\leq R/8$, $d({  y}_i, {  y}_0)\leq
%R/8$, and $3R/8< d({  x}_j, {  x}_0)\leq R/2$, $3R/8<d({
%y}_j, {  y}_0)\leq R/2$.
For these sample points $i$ and $j$,
\begin{equation}\label{app-proofctdeq1}
3R/8\leq d({  x}_j, {  x}_0)\leq d({  x}_j, {  x}_i)+d({  x}_i, {
x}_0) \leq d({  x}_j, {  x}_i)+R/8
\end{equation}
where the second inequality is the triangle inequality, and the
first and third inequalities follow since $(x_j,y_j)\in
\mathcal{A}_2$ and  $(x_i,y_i)\in \mathcal{A}_1$.
 It follows from (\ref{app-proofctdeq1}) that
\begin{equation}\label{app-proofctdeq11}
d({  x}_i, {  x}_j)\geq R/4, \quad d({  y}_i, {  y}_j)\geq R/4.
\end{equation}

Moreover, if a sample point $k$ is closer to $i$ than to $j$ both in
the $  X$ vector and in the $  Y$ vector, then it is within the $ x$
and $  y$ spheres of radius $R$:
\begin{lemma}\label{app-proofctdlemma}
If $d({  x}_k,{  x}_i)<d({  x}_i,{  x}_j)$, then $d({  x}_k,{
x}_0)\leq R$. Similarly, if $d({  y}_k,{  y}_i)<d({  y}_i,{ y}_j)$,
then $d({  y}_k,{  y}_0)\leq R$.
\end{lemma}
%\begin{proof}
Proof: Since the proof follows the same steps for ${  x}_k$ and ${
y}_k$, we only show it for the $  x$ coordinates. The result follows
by applying the triangle inequality several times,

\begin{eqnarray}
 d({  x}_k, {  x}_0) && \leq d({  x}_k, {  x}_i)+d({
x}_i, {  x}_0) \leq d({  x}_j, {  x}_i) +  d({  x}_i, { x}_0)
\nonumber
\\
&& \leq d({  x}_j, {  x}_0) +  2 d({  x}_i, {  x}_0) \nonumber \leq
R/2 + 2R/8 = 6R/8\leq R.
\end{eqnarray}
%\end{proof}

The consequence of Lemma \ref{app-proofctdlemma} is that for all
such samples $k$,  $({  x}_k, {  y}_k)\in \mathcal{A}$.
%$h({  x}_k,{  y}_k) -  f({  x}_k)g({  y}_k)\geq c$.

Moreover, all points that are within the $  x$ and $  y$ spheres of
radius $R/8$ are closer to $i$ than the point $j$:
\begin{lemma}\label{app-proofctdlemma2}
If $d({  x}_k,{  x}_0)<R/8$, then $d({  x}_k,{  x}_i)<d({ x}_i,{
x}_j)$. Similarly, if $d({  y}_k,{  y}_0)<R/8$, then $d({ y}_k,{
y}_i)<d({  y}_i,{  y}_j)$.
\end{lemma}
%\begin{proof}
Proof:  Since the proof follows the same steps for ${  x}_k$ and ${
y}_k$, we only show it for the $  x$ coordinates. Applying the
triangle inequality, $ d({  x}_k, {  x}_i)\leq d({  x}_k, {
x}_0)+d({  x}_i, { x}_0) \leq R/8+R/8 = R/4. $ The result follows
from (\ref{app-proofctdeq11}).
%\end{proof}
Therefore, if $(x_k,y_k)\in \mathcal{A}_1$, then $k$ is closer to
$i$ than to $j$ in both $X$ and $Y$.

 By the law of large numbers, almost surely
\begin{equation} \label{app-proofctdeq4}
\lim_{N\rightarrow \infty} \frac{A_{11}(i,j)}{N-2} = p\{{ x}_0({
u}),{ y}_0({  s}) \} \int_{\mathcal{A}_3} h\{{  x}({  v}),{ y}({
t})\}d{ x}({  v})d{  y}({  t})
\end{equation}
\begin{equation} \label{app-proofctdeq5}
\lim_{N\rightarrow \infty} \frac{A_{1\cdot}(i,j)}{N-2} = p\{{ x}_0({
u})\} \int_{\mathcal{A}_4}f\{{  x}({  v})\}d{  x}({  v})
\end{equation}
\begin{equation} \label{app-proofctdeq6}
\lim_{N\rightarrow \infty} \frac{A_{\cdot1}(i,j)}{N-2}
 =  p\{{
y}_0({  s})\}\int_{\mathcal{A}_5}g\{{  y}({  t})\}d{  y({  t})}
\end{equation}
where  $\mathcal{A}_3 =  \{({  x}, {  y}): d({  x},{  x}_i)< d({
x}_i,{  x}_j), d({  y},{  y}_i)<d({  y}_i,{  y}_j) \},$
$\mathcal{A}_4 = \{{  x}: d({  x},{  x}_i)< d({  x}_i,{  x}_j) \}$,
and $\mathcal{A}_5 = \{{  y}: d({  y},{  y}_i)<d({  y}_i,{  y}_j)
\}$ .

Recall that $S(i,j) =\sum_{k=1}^2\sum_{l=1}^2{\{A_{k,l}(i,j) -
A_{k\cdot}(i,j)A_{\cdot l}(i,j)/(N-2)\}^2}/\{A_{k\cdot}(i,j)A_{\cdot
l}(i,j)/(N-2)\}$. It is enough to look at the term with $l=1$ and
$k=1$ in $S(i,j)$, i.e. the term $$S_1(i,j) = \frac{\{A_{11}(i,j) -
A_{1\cdot}(i,j)A_{\cdot
1}(i,j)/(N-2)\}^2}{A_{1\cdot}(i,j)A_{\cdot1}(i,j)/(N-2)}.$$ It
follows that $S(i,j)\geq S_1(i,j)$, and therefore that our test
statistic $T \geq \sum_{i=1}^N \sum^N_{\substack{
 j\neq i\\ j=1} }S_1(i,j). $

By Slutzky's theorem and the continuous mapping theorem, almost
surely
\begin{eqnarray}\label{app-proofctdeq7}
 \lim_{N\rightarrow \infty} \frac{S_1(i,j)}{N-2} &=&
\lim_{N\rightarrow \infty}  \frac{1}{N-2}\frac{\{A_{11}(i,j) -
A_{1\cdot}(i,j)A_{\cdot1}(i,j)/(N-2)\}^2}{A_{1\cdot}(i,j)A_{\cdot1}(i,j)/(N-2)}
\nonumber \\
&  =& \frac{(\int_{\mathcal{A}_3} [p\{{  x}_0({  u}),{  y}_0({ s})
\} h\{{  x}({  v}),{  y}({  t})\} - p\{{  x}_0({  u})\} f\{{  x}({
v})\} p\{{ y}_0({  s})\}g\{{  y}({  t})\} ]d{  x}({  v})d{  y}({
t}))^2}{\int_{\mathcal{A}_3} [ p\{{  x}_0({  u})\} f\{{  x}({ v})\}
p\{{ y}_0({  s})\}g\{{  y}({  t})\} ]d{  x}({  v})d{  y}({  t})}.
\end{eqnarray}
We shall show that this limit can be bound from below by a positive
constant that depends on $({  x}_0, {  y}_0)$ but not on $i$ and
$j$. From Lemma \ref{app-proofctdlemma} it follows that
${\mathcal{A}_3}\subseteq {\mathcal{A}}$, and from Lemma
\ref{app-proofctdlemma2} it follows that  ${\mathcal{A}_1}\subseteq
{\mathcal{A}_3}$, and therefore  a positive lower bound on the
numerator of (\ref{app-proofctdeq7}) can be obtained:
\begin{eqnarray}
&& \int_{\mathcal{A}_3} [p\{{  x}_0({  u}),{  y}_0({  s}) \} h\{{
x}({ v}),{  y}({  t})\} - p\{{  x}_0({  u})\} f\{{  x}({  v})\} p\{{
y}_0({ s})\}g\{{  y}({  t})\} ]d{  x}({  v})d{  y}({  t})
\nonumber \\
 &&\geq c
\int_{\mathcal{A}_3} d{  x}({  v})d{  y}({  t}) \geq c
\int_{\mathcal{A}_1} d{  x}({  v})d{  y}({  t}) \nonumber.
\end{eqnarray}
Moreover, $\int_{\mathcal{A}_3} \{ p\{{  x}_0({  u})\} f\{{  x}({
v})\} p\{{ y}_0({  s})\}g\{{  y}({  t})\} \}d{ x}({  v})d{  y}({ t})
\leq 1$.  Therefore, denoting the lower bound by $c'= \{ c
\int_{\mathcal{A}_1} d{  x}({  v})d{  y}({  t})\}^2$, it follows
that $S_1(i,j)/(N-2)$ converges almost surely to a constant larger
than $c'>0$. Therefore, $S_1(i,j)>(N-2)c'/2$ with probability going
to 1 as $N\rightarrow \infty$. Since, moreover, the number of pairs
of points $i$ and $j$ such that $(x_i,y_i)\in \mathcal{A}_1$ and
$(x_j,y_j) \in \mathcal{A}_2$, divided by $f_1 f_2 N^2$, converges
almost surely to 1, it follows that there exists a constant $\delta$
such that $\lim_{N\rightarrow \infty} Pr(T>\delta N^3) =1 $.

Under the null hypothesis, for large enough sample size $N$,
$S(i,j)$ is distributed $\chi^2$ with 1 degree of freedom.
Therefore, the null expectation of $T$ is approximately $N(N-1)$,
and the null variance is bounded above by a term of order $N^4$
(more precisely, by $\{N(N-1)\}^2 2$). Since $\sum_{i=1}^N
\sum^N_{\substack{
 j=1 \\ j\neq i} }S(i,j)$
is of order of magnitude of $N^3$, it follows that $T$ will be
rejected with probability 1.

%\bibliographystyle{apalike}
%\bibliography{references}

\section{ Supplementary Material }

\subsection{Proofs} The proof of case 1 is omitted, since it is very
similar to the more complex case 2. The proofs of the countable case
3, and the mixed case where one random vector is discrete and the
other continuous, are given, respectively, in Sections
\ref{subsec-proofcase3} and \ref{subsec-proofcase4} below.

\subsubsection{Proof of the countable case
3}\label{subsec-proofcase3} Suppose $X\in \Re^p$ and $Y\in \Re^q$
are both discrete with countable support. $H_0$ is false implies
that there exists at least one pair of atoms $(x_0, y_0)$ such that
$Pr(X=x_0, Y=y_0)> Pr(X=x_0)Pr(Y=y_0)$. We expect $N Pr(X=x_0,
Y=y_0)$ points to have values $(x_0,y_0)$.  Let $i$ and $j$ be two
such points.
 By the law of large numbers, almost surely
\begin{equation}
\lim_{N\rightarrow \infty} \frac{A_{11}(i,j)}{N-2} = Pr(X = x_0, Y =
y_0),\quad \lim_{N\rightarrow \infty} \frac{A_{1\cdot}(i,j)}{N-2}
=Pr(X = x_0), \quad \lim_{N\rightarrow \infty}
\frac{A_{\cdot1}(i,j)}{N-2} = Pr(Y=y_0). \nonumber
\end{equation}

Recall that $$S(i,j) =\sum_{k=1}^2\sum_{l=1}^2{\{A_{k,l}(i,j) -
A_{k\cdot}(i,j)A_{\cdot l}(i,j)/(N-2)\}^2}/\{A_{k\cdot}(i,j)A_{\cdot
l}(i,j)/(N-2)\}.$$ It is enough to look at the term with $l=1$ and
$k=1$ in $S(i,j)$, i.e. the term $$S_1(i,j) = \frac{\{A_{11}(i,j) -
A_{1\cdot}(i,j)A_{\cdot
1}(i,j)/(N-2)\}^2}{A_{1\cdot}(i,j)A_{\cdot1}(i,j)/(N-2)}.$$ It
follows that $S(i,j)\geq S_1(i,j)$, and therefore that our test
statistic $T \geq \sum_{i=1}^N \sum^N_{\substack{
 j\neq i\\ j=1} }S_1(i,j). $

By Slutzky's theorem, almost surely
\begin{eqnarray}
 \lim_{N\rightarrow \infty} \frac{S_1(i,j)}{N-2} &=&
\lim_{N\rightarrow \infty}  \frac{1}{N-2}\frac{\{A_{11}(i,j) -
A_{1\cdot}(i,j)A_{\cdot1}(i,j)/(N-2)\}^2}{A_{1\cdot}(i,j)A_{\cdot1}(i,j)/(N-2)}
\nonumber \\
& = & \frac{\{Pr(X = x_0, Y = y_0)  - Pr(X=x_0)
Pr(Y=y_0)\}^2}{Pr(X=x_0) Pr(Y=y_0)}. \nonumber
\end{eqnarray}

It follows that $S_1(i,j)/(N-2)$ converges almost surely to a
positive constant $c'>0$. Therefore, $S_1(i,j)>(N-2)c'/2$ with
probability going to 1 as $N\rightarrow \infty$. Since we have order
of magnitude of $N^2$ pairs of points $i$ and $j$ that satisfy the
inequality $S_1(i,j)>(N-2)c'/2$, it follows that there exists a
constant $\delta$ such that $\lim_{N\rightarrow \infty} Pr(T>\delta
N^3) =1 $.  By the same argument as in the last paragraph of the
Appendix in the main text, it therefore follows that $T$ will be
rejected with probability 1.

\subsubsection{Proof of mixed case 4}\label{subsec-proofcase4}
Suppose $X\in \Re^p$ is discrete with countable support, and $Y\in
\Re^q$ has a continuous density given $X$, denoted by $h(y \mid
X=x)$, and a marginal density $g(y)$. $H_0$ is false implies that
there exists at least one pair of points $x_0, y_0$ such that
$Pr(X=x_0)h(Y=y_0 \mid X=x_0)> Pr(X=x_0)g(Y=y_0)$. Since $h(\cdot
\mid X=x_0)$ is continuous, there exists a radius $R$ such that
$Pr(X=x_0)h(Y=y \mid X=x_0)> Pr(X=x_0)g(Y=y)$ for $(x,y) \in
\mathcal{A} = \{(x,y): x = x_0, d(y,y_0)<R \}$. The set
$\mathcal{A}$ has positive probability, and moreover
$$\min_{\mathcal{A}}\{Pr(X=x_0)h(Y=y \mid X=x_0) -
Pr(X=x_0)g(Y=y)\}>0. $$ Denote this minimum by the positive constant
$c$.

Clearly the following two subsets of $\mathcal{A}$ have positive
probability as well: $$\mathcal{A}_1 = \{({  x}, {  y}): x=x_0, d({
y},{ y}_0)<R/8 \}$$ and
$$\mathcal{A}_2 = \{({  x}, {  y}): x=x_0, 3R/8<d({  y},{  y}_0)<R/2 \}.$$ Denote the
probabilities of $\mathcal{A}_1$ and $\mathcal{A}_2$ by $f_1$ and
$f_2$ respectively. Therefore, we expect $(N f_1) (N f_2)$ pairs of
sample points $i$ and $j$ such that $({  x}_i, {  y}_i) \in
\mathcal{A}_1$ and $({ x}_j, {  y}_j) \in \mathcal{A}_2$.

For these sample points $i$ and $j$, $d(y_i, y_j)\geq R/4$. From
Lemma 1 in the Appendix, if $d({  y}_k,{ y}_i)<d({ y}_i,{ y}_j)$,
then $d({ y}_k,{  y}_0)\leq R$. From Lemma 2 in the Appendix,  if
$d({  y}_k,{  y}_0)<R/8$, then $d({ y}_k,{ y}_i)<d({  y}_i,{ y}_j)$.
Therefore, if $(x_k,y_k)\in \mathcal{A}_1$, then $k$ is closer to
$i$ than to $j$ in $Y$.

By the law of large numbers, almost surely

\begin{equation} \label{app-proofctdeq4}
\lim_{N\rightarrow \infty} \frac{A_{11}(i,j)}{N-2} = Pr(X=x_0)
\int_{\mathcal{A}_3} h( y \mid X=x_0) d{ y}
\end{equation}
\begin{equation} \label{app-proofctdeq5}
\lim_{N\rightarrow \infty} \frac{A_{1\cdot}(i,j)}{N-2} = Pr(X=x_0)
\end{equation}
\begin{equation} \label{app-proofctdeq6}
\lim_{N\rightarrow \infty} \frac{A_{\cdot1}(i,j)}{N-2}
 =  \int_{\mathcal{A}_4}g(y) dy
\end{equation}
where  $\mathcal{A}_3 =  \{({  x}, {  y}): x=x_0, d({  y},{
y}_i)<d({ y}_i,{  y}_j) \},$ , and $\mathcal{A}_4 = \{{  y}: d({
y},{ y}_i)<d({ y}_i,{  y}_j) \}$ .

Recall that $$S(i,j) =\sum_{k=1}^2\sum_{l=1}^2{\{A_{k,l}(i,j) -
A_{k\cdot}(i,j)A_{\cdot l}(i,j)/(N-2)\}^2}/\{A_{k\cdot}(i,j)A_{\cdot
l}(i,j)/(N-2)\}.$$ It is enough to look at the term with $l=1$ and
$k=1$ in $S(i,j)$, i.e. the term $$S_1(i,j) = \frac{\{A_{11}(i,j) -
A_{1\cdot}(i,j)A_{\cdot
1}(i,j)/(N-2)\}^2}{A_{1\cdot}(i,j)A_{\cdot1}(i,j)/(N-2)}.$$ It
follows that $S(i,j)\geq S_1(i,j)$, and therefore that our test
statistic $T \geq \sum_{i=1}^N \sum^N_{\substack{
 j\neq i\\ j=1} }S_1(i,j). $

By Slutzky's theorem and the continuous mapping theorem, almost
surely
\begin{eqnarray}
 \lim_{N\rightarrow \infty} \frac{S_1(i,j)}{N-2} &=&
\lim_{N\rightarrow \infty}  \frac{1}{N-2}\frac{\{A_{11}(i,j) -
A_{1\cdot}(i,j)A_{\cdot1}(i,j)/(N-2)\}^2}{A_{1\cdot}(i,j)A_{\cdot1}(i,j)/(N-2)}
\nonumber \\
&  =& \frac{Pr(X = x_0)[\int_{\mathcal{A}_3}  \{ h( y \mid X=x_0) d{
y} - g(y)\}dy]^2}{\int_{\mathcal{A}_4} g(y)dy} \nonumber
\end{eqnarray}

It follows that $S_1(i,j)/(N-2)$ converges almost surely to a
positive constant $c'>0$. Therefore, $S_1(i,j)>(N-2)c'/2$ with
probability going to 1 as $N\rightarrow \infty$. Since we expect $(N
f_1) (N f_2)$ pairs of sample points $i$ and $j$  that satisfy the
inequality $S_1(i,j)>(N-2)c'/2$, it follows that there exists a
constant $\delta$ such that $\lim_{N\rightarrow \infty} Pr(T>\delta
N^3) =1 $.  By the same argument as in the last paragraph in the
Appendix of the main text, it therefore follows that $T$ will be
rejected with probability 1.

\subsection{Computational Complexity}\label{subsec-comp} In this
Section we give a $C$ implementation of the computation of $\{
inv(j): j=1,\ldots,N, j\neq i \}$ in  order of magnitude $N\log N$.
The algorithm uses an adaptation of the classic merge sort
algorithm. The basic idea is to split the array in half and sort
each half while counting the number of inversions for each element
in each half. In the merging stage of both halves, if an element in
the right side is smaller than an element in the left side, it means
that the number of inversions for the smaller element should be
updated by adding to it the number of elements on the left side
which are larger than it. The complexity of this algorithm $T(N)$
respects the recursion $T(N)=2T(N/2) +O(N)$ and therefore it is
$T(N)=O(N\log N)$. The C code is given below. {\small
\begin{verbatim}

int Inversions(int *permutation, int *source, int
*inversion_count,int dim) {
    if (dim==1)
        return 0;
    else{
        Inversions(permutation, source, inversion_count, dim/2);
        Inversions(&permutation[dim/2], &source[dim/2], inversion_count,dim/2);
        Merge(permutation, source, inversion_count, dim);
    }
    return 0;
}

int Merge(int *permutation, int *source, int *inversion_count, int
dim) {
    int i;
    int left[MAX_DIM], right[MAX_DIM], left_source[MAX_DIM], right_source[MAX_DIM];
    int left_index=0, right_index=0;
    for (i=0;i<dim/2;i++){
        left[i]=permutation[i];
        left_source[i]=source[i];
    }
    for(i=0;i<dim/2;i++){
        right[i]=permutation[i+dim/2];
        right_source[i]=source[i+dim/2];
    }
    for(i=0;i<dim;i++){
        if ( (left_index<dim/2) && (right_index<dim/2)){
             if (left[left_index]<right[right_index]){
                permutation[i]=left[left_index];
                source[i]=left_source[left_index];
                left_index++;
            }
            else{
                permutation[i]=right[right_index];
                source[i]=right_source[right_index];
                printf("adding %d invs to %d\n", dim/2-left_index, source[i]);
                inversion_count[source[i]]+=(dim/2-left_index);
                right_index++;
            }
        }
        else{
            if (left_index<dim/2){
                permutation[i]=left[left_index];
                source[i]=left_source[left_index];
                left_index++;
            }
            if (right_index<dim/2){
                permutation[i]=right[right_index];
                source[i]=right_source[right_index];
                right_index++;
            }

        }
    }
    return 0;
}

\end{verbatim}
}

\subsection{Simulations} 
\begin{figure}[!tpb]
\centering
\includegraphics[width=4.0cm, height=4.0cm]{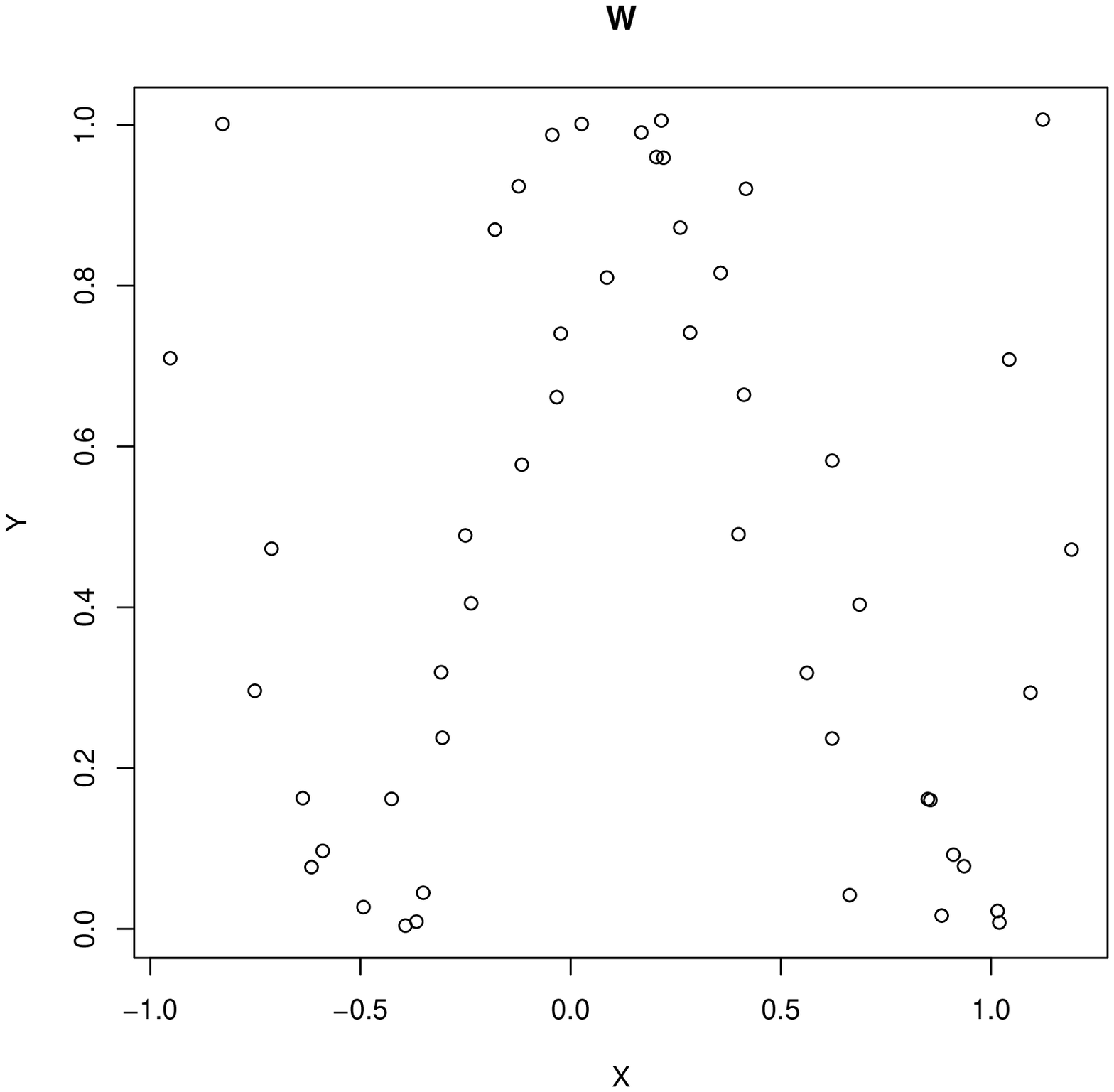}
\includegraphics[width=4.0cm, height=4.0cm]{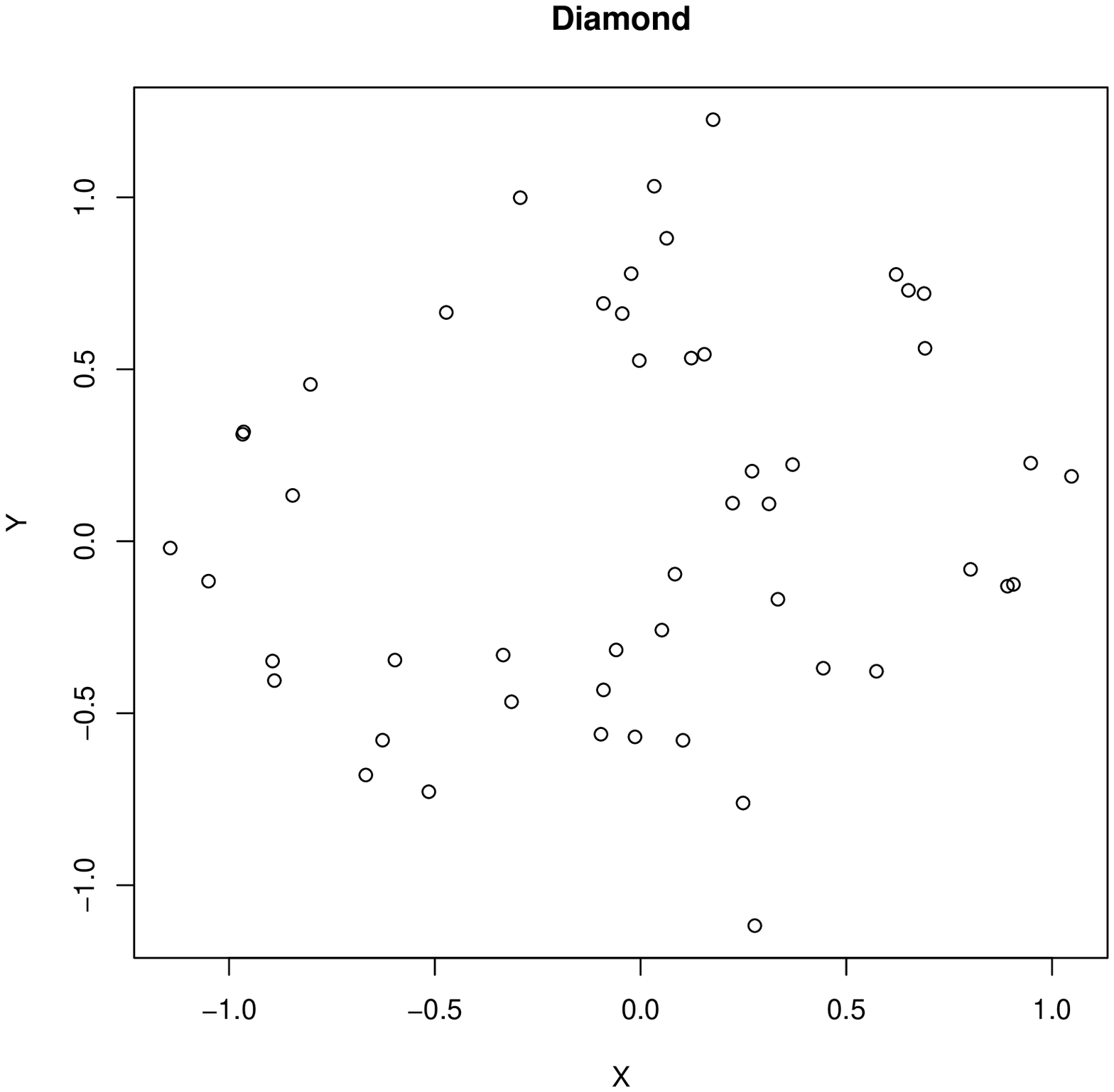}
\includegraphics[width=4.0cm, height=4.0cm]{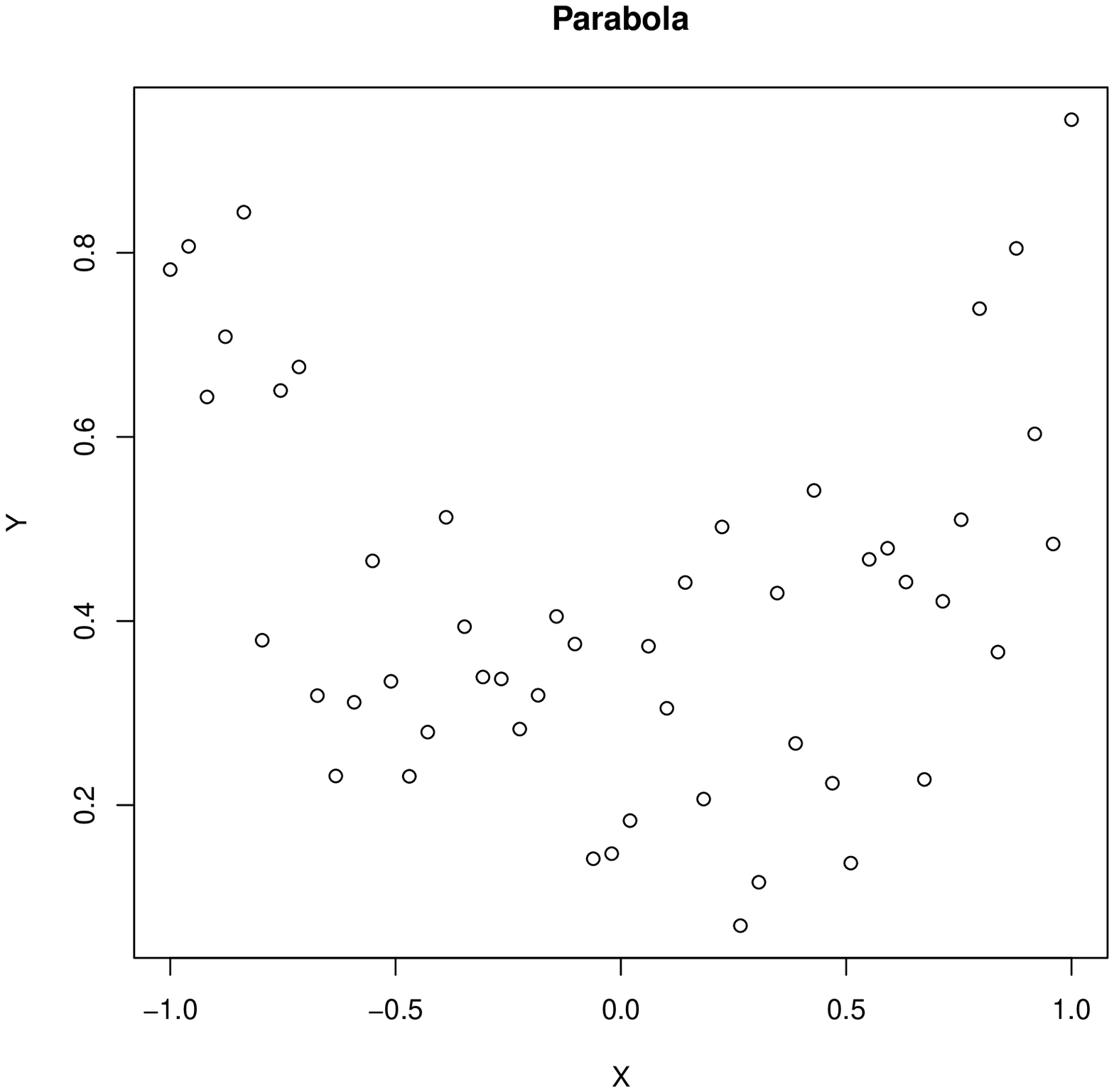}
\includegraphics[width=4.0cm, height=4.0cm]{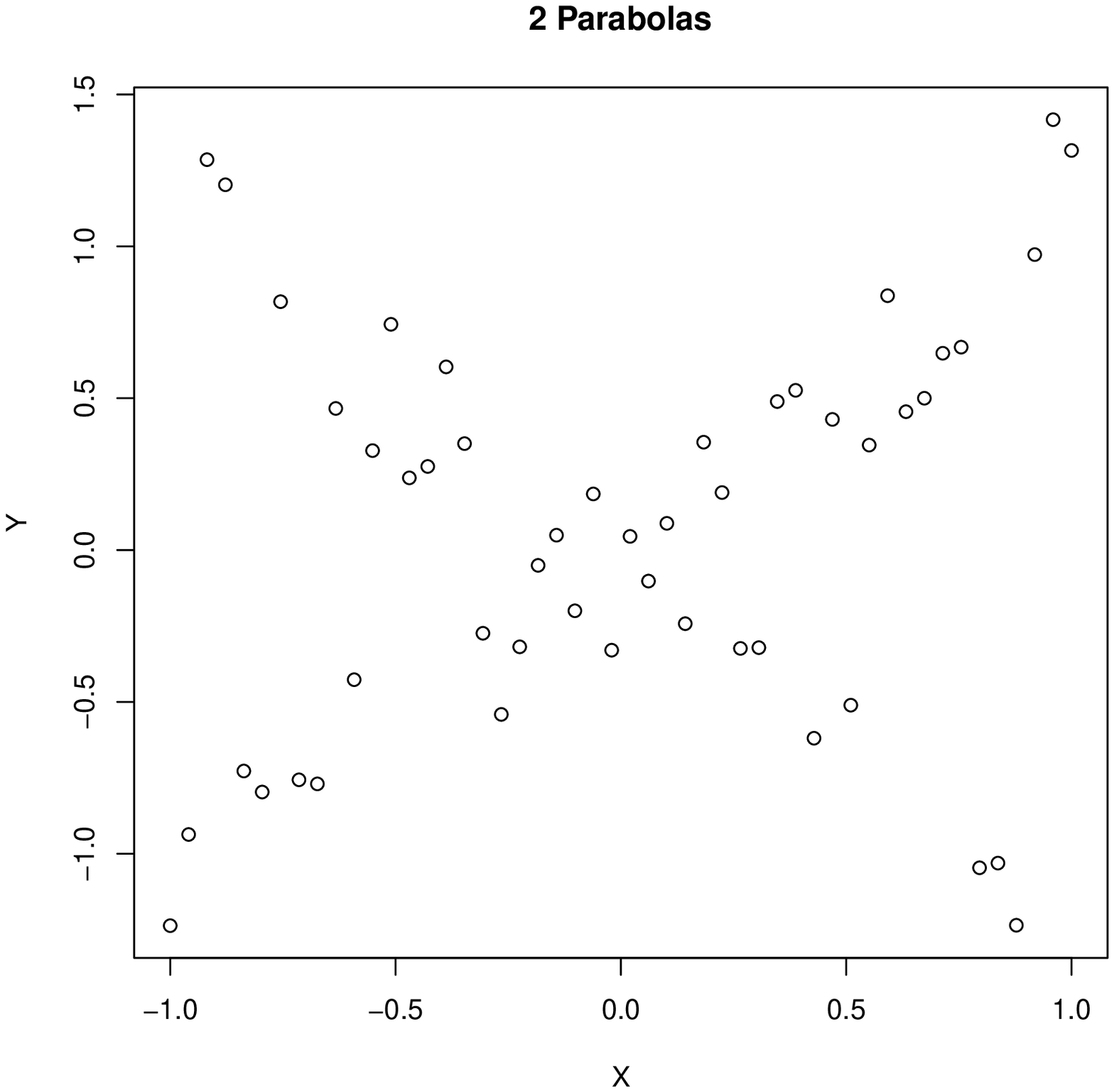}
\includegraphics[width=4.0cm, height=4.0cm]{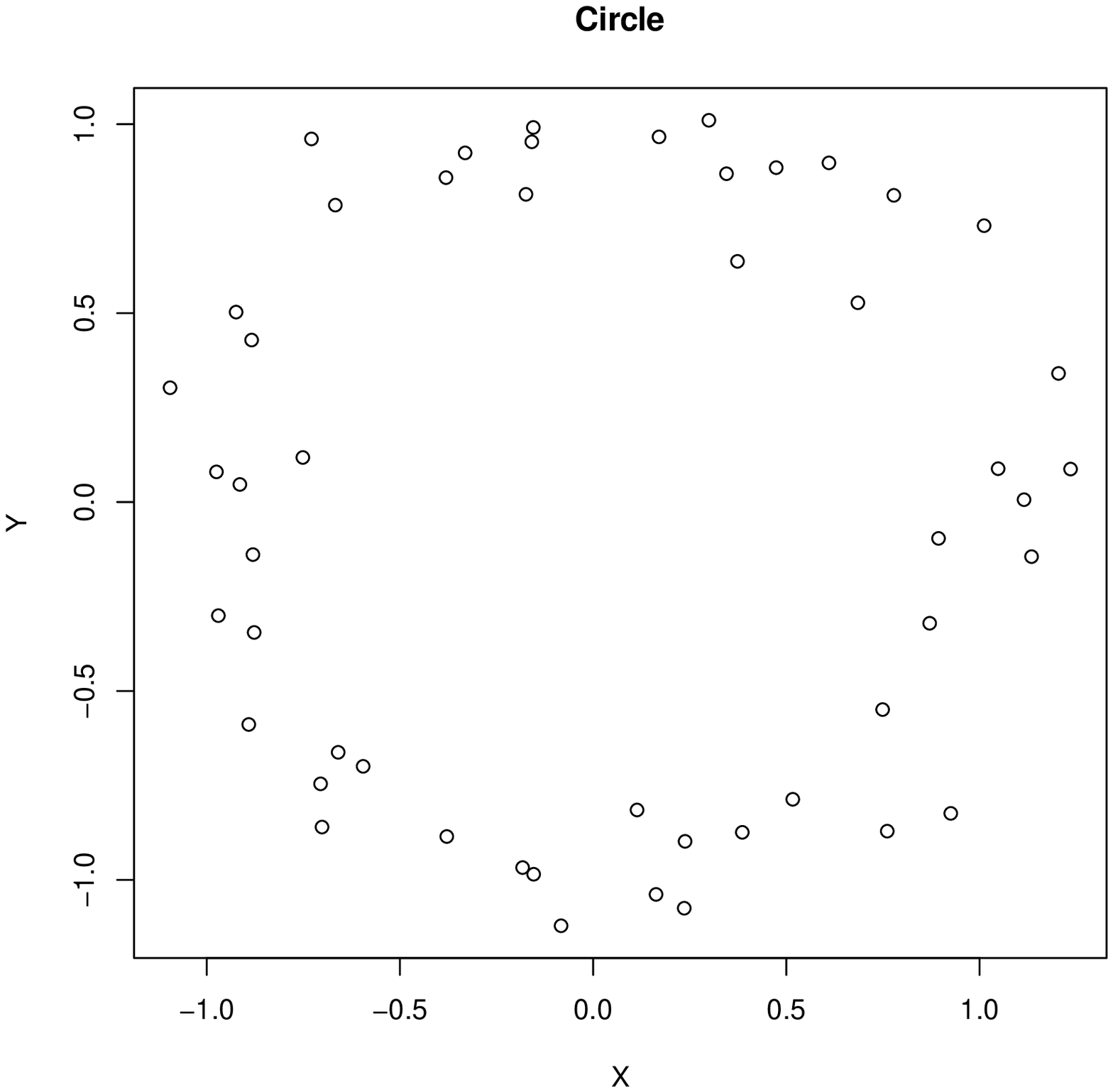}
\includegraphics[width=4.0cm, height=4.0cm]{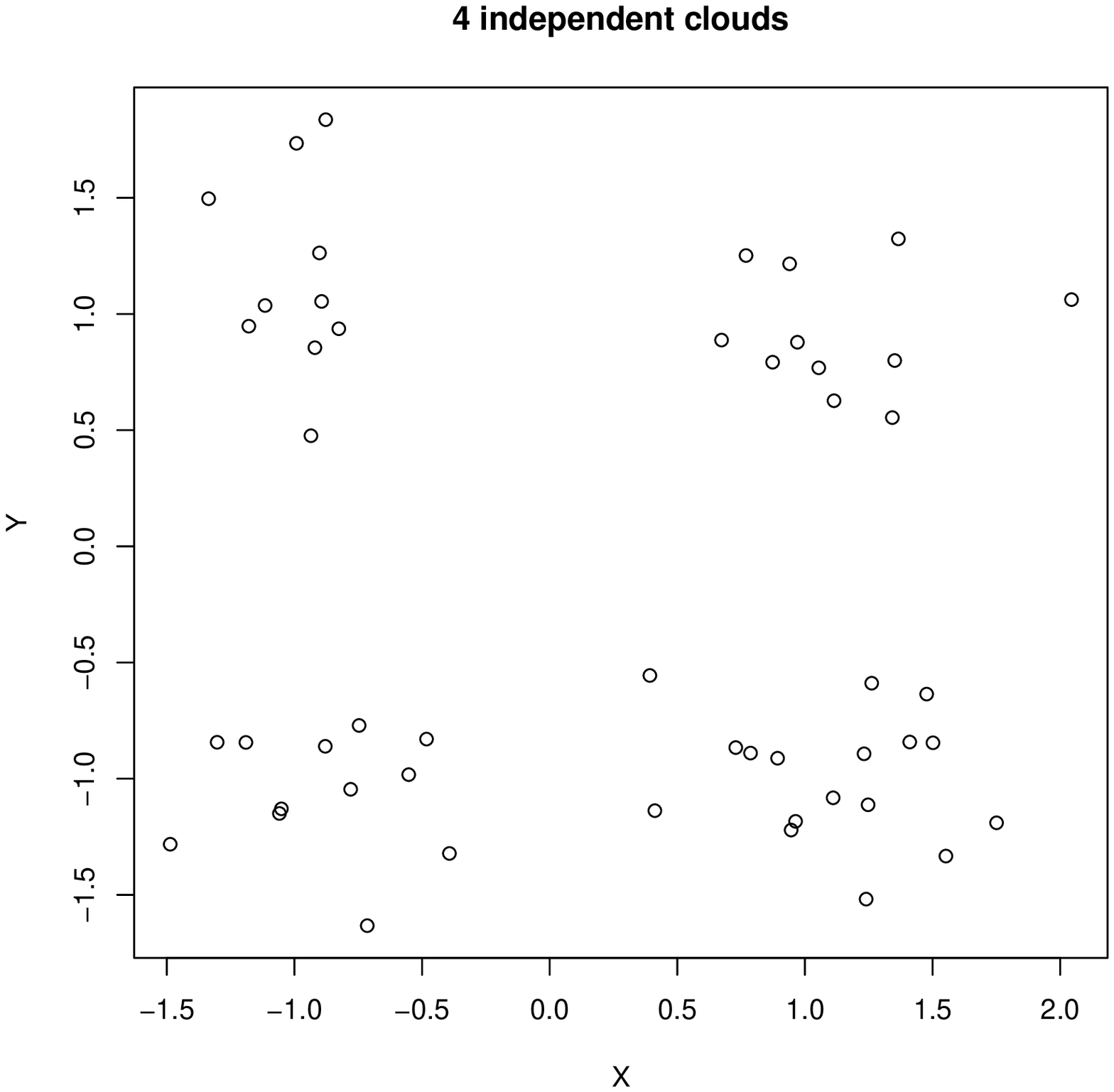}
\caption{Six simulated examples of unusual bivariate distributions;
a sample of size N=50 from each distribution. }\label{fig-newton}
\end{figure}

In the simulations presented in the main
text, we first considered the six simulated examples of unusual
bivariate distributions. Figure \ref{fig-newton} shows the scatter
plots for a sample of size $N=50$ from each of these distributions.

\begin{figure}[!tpb]
\centering
\includegraphics[width=4.0cm, height=4.0cm]{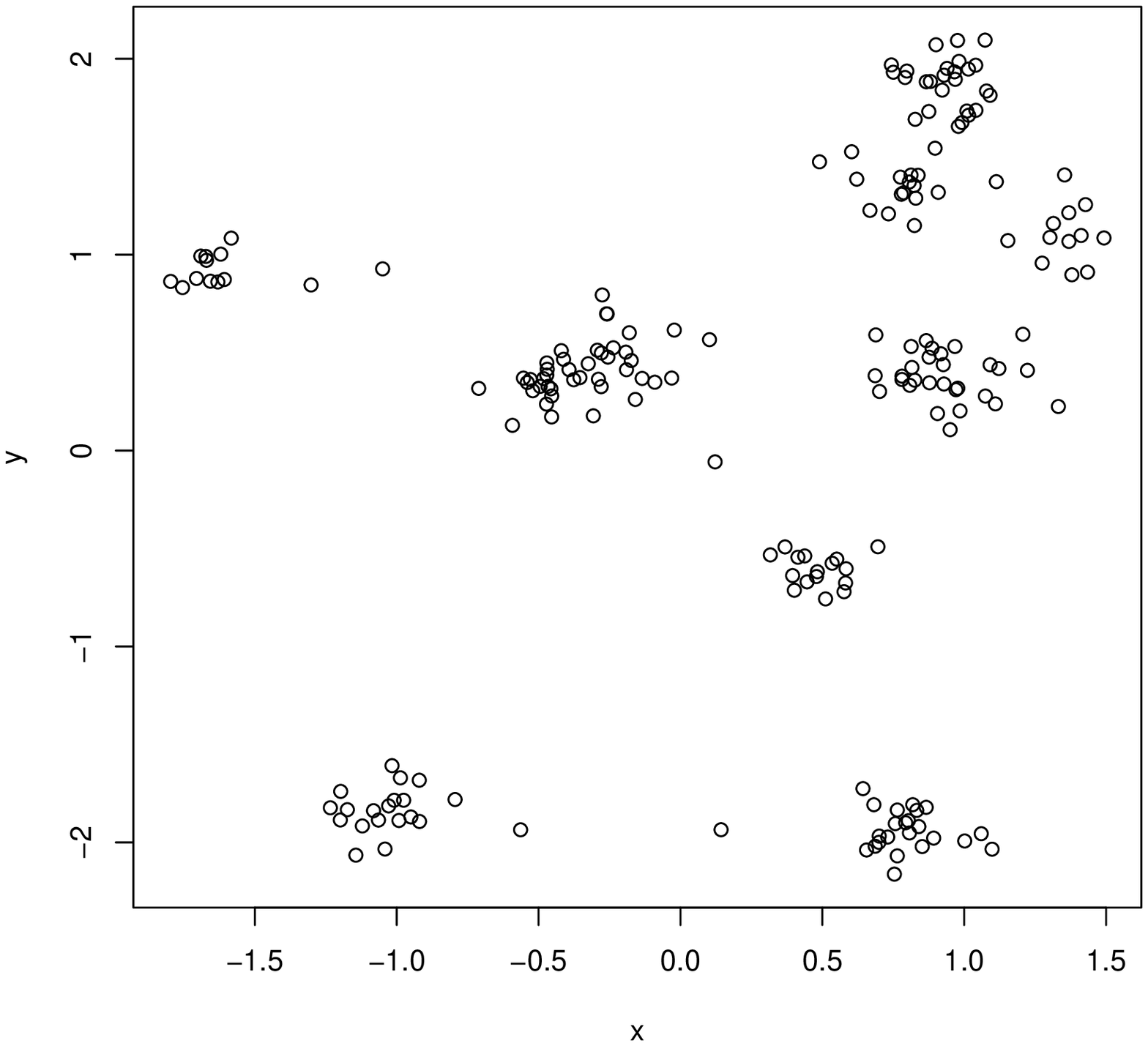}
\includegraphics[width=4.0cm, height=4.0cm]{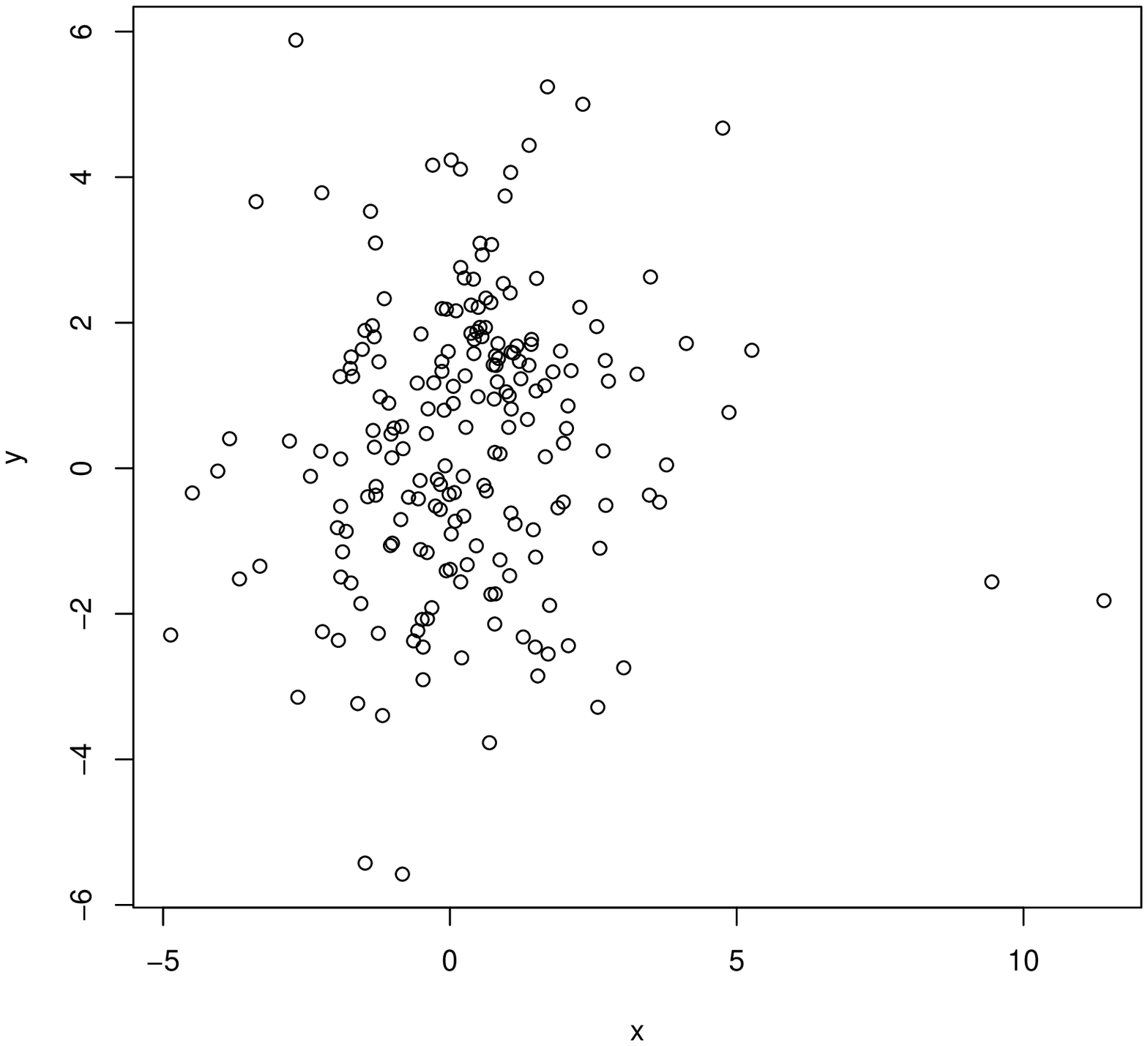}
\caption{A scatter plot of the first coordinate in the mixture
distribution of 10 components, where each coordinate has a t
distribution with 3df around a different center. Left panel, noise
10 times smaller than generated; Right panel, noise used in the
simulation.}\label{fig-mixed}
\end{figure}

In the simulations presented in the main text, the last example was
of a mixture distribution in 1000 dimensions. Figure \ref{fig-mixed}
shows the first coordinate of $X$ and $Y$ in a setting where the
standard deviation of the noise is 10 times smaller than actually
generated (Left), as well as with the actual noise used in the
simulation (Right panel), for the multivariate $t$ distribution with
3df.

A more sophisticated scenario in 100 dimensions, which includes both
a monotone and non-monotone component, is the following: $Y_j =
\beta_1 X_j+\beta_2 X_j^2 + \epsilon_j, j\in I_1$ and $Y_j =
\epsilon_j, j \in
 \{1,\ldots,100\}\backslash I_1$, with $\epsilon_j \sim N(0, \Sigma_X)$ and $X \sim
N(0,\Sigma_X)$. The covariance matrix $\Sigma_X$ is block diagonal,
with symmetric correlation of 0.9 in the first block, 0.8 in the
second block, etc. The last block has 0 correlation, and the
diagonal entries of $\Sigma_X$ are 1. In the null setting where $I_1
= \emptyset$, the empirical power for the new test, based on 1000
simulations, was 0.046, 0.043, and 0.051 for $N=30,40$, and $50$,
respectively. Table \ref{tab-referee} shows the power of a test at
level $0.05$ for dCov as well as for the new test for $\beta_1 = 1,
\beta_2=4, \sigma^2 = 9$, and two configurations of $I_1$. The power
of the new test is better than that of dCov in the settings
considered, in which the non-monotone part of the relationship has a
stronger effect than the monotone part of the relationship.
Moreover, the power of both tests is larger in the first setting, of
strong dependence between the coordinates of $X$, than in the second
setting, where the dependence across coordinates is weaker, since in
the first setting the highly associated components of $X$ cause
dependence between each coordinate of $Y$ with several coordinates
of $X$.

\begin{table}[!t]
\caption{The power of a test at level $0.05$ per sample size from a
100 dimensional joint distribution, where $Y_j =  X_j+ 4 X_j^2 +
\epsilon_j, j\in I_1$ and $Y_j = \epsilon_j, j\in \{1,\ldots,100\}
\backslash I_1$, with $\epsilon_j \sim N(0, 9)$. The results are
based on 1000 simulations.}\label{tab-referee}
\begin{center}
\begin{tabular}{r|r|r|r}
  $I_1$&  Sample size & dCov  & new test\\ \hline
$\{1,\ldots, 10, 51, \ldots, 55\}$ & $N=30$ & 0.382 & 0.629 \\
& $N=40$ & 0.456 & 0.782 \\
& $N=50$ & 0.541 & 0.879 \\
 $ \{41, \ldots, 50, 91, \ldots, 100\}$ & $N=30$ & 0.246 &0.243\\
& $N=40$ &0.271 &0.340\\
& $N=50$ &0.293 &0.474\\
& $N=60$ &0.359 &0.553\\
& $N=70$ &0.369 &0.626\\
& $N=80$ &0.433 &0.673\\
 \hline
\end{tabular}
\end{center}
\end{table}

\subsection{A univariate example}\label{SM-sec-example}
\begin{figure}[!tpb]
\centering
\includegraphics[width=7cm, height=7cm]{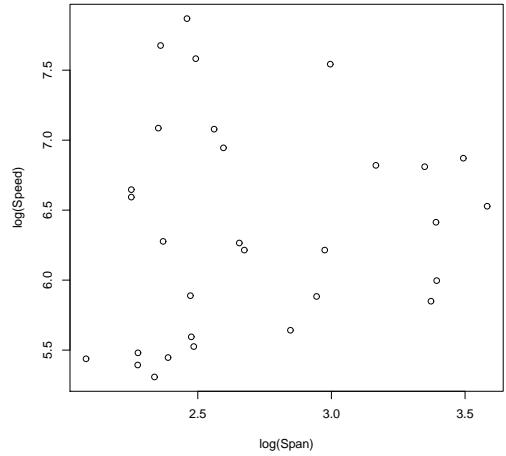}
\caption{The scatter of wing span vs. speed on the log scale for a
sample of 30 points. The $p$-value from the $dCov$ test and our
proposed test were 0.21 and 0.03, respectively.}\label{figAircraft}
\end{figure}

\begin{figure}[!tpb]
\centering
\includegraphics[width=7cm, height=7cm]{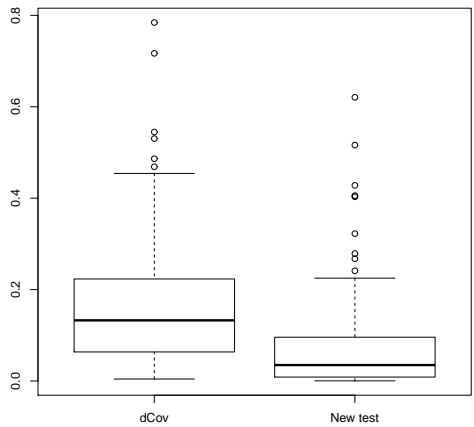}
\caption{The boxplots of the 100 $p$-values for dCov and the
proposed test based on a random sample of 30 points from the
Aircraft data.}\label{fig}
\end{figure}

\cite{Szekely09} examined the Saviotti aircraft data  of
\cite{Saviotti96}, that records six characteristics of aircraft
designs  during the twentieth century. They consider two variables,
wing span (m) and speed (km/h) for the 230 designs of the third (of
three) periods. This example and the data (aircraft) are from
\cite{Bowman97}.  They showed that the dCov test of independence of
log(Speed) and log(Span) in period 3 is significant (p-value $\leq
0.00001$), while the Pearson correlation test is not significant
(p-value = 0.8001). Our proposed test is also highly significant
(p-value $\leq 0.00001$). Moreover, if we take a random sample of 30
observations and apply the dCov test and the proposed test to this
small random sample, then we typically get smaller $p$-values using
our proposed test than using the $dCov$ test. Specifically,
repeating the testing of a random sample of 30 observations 100
times, the p-value of our proposed test was below 0.05 for 58/100
simulation runs, whereas for dCov only for 18/100 simulation runs.
Figure \ref{figAircraft} shows the scatter plot of wing span vs.
speed on the log scale for a sample of 30 points. The relationship
appears fan-like. For this particular sample, the $p$-value from the
$dCov$ test and our proposed test were 0.21 and 0.03, respectively.
Figure \ref{fig} shows the distribution of the 100 $p$-values for
each of the tests.

%\bibliographystyle{apalike}
%\bibliography{references}

\end{document}